\documentclass[aps,nofootinbib,superscriptaddress,tightenlines,twocolumn,pra]{revtex4-1}

\usepackage{amsmath,amssymb,amstext,amsthm}
\usepackage[pdftex]{graphicx}
\usepackage{braket}
\usepackage{bbm}
\usepackage{xcolor}
\definecolor{dred}{rgb}{.8,0.2,.2}
\definecolor{ddred}{rgb}{.8,0.5,.5}
\definecolor{dblue}{rgb}{.2,0.2,.8}

\usepackage[pdftex,letterpaper=true,pagebackref=false]{hyperref}
\hypersetup{
pdftitle={Quantum Information Approach to Dynamical Diffraction Theory},
pdfauthor={J. Nsofini},
pdfsubject={Quantum information, quantum physics, neutron interferometry}
	pdfnewwindow=true, 
	colorlinks=true, 
	linkcolor=blue, 
	citecolor=blue, 
	filecolor=blue, 
	urlcolor=blue
}

\def\beq{\begin{equation}}
\def\eeq{\end{equation}}
\def\bsp{\begin{split}}
\def\esp{\end{split}}
\def\bea{\begin{eqnarray}}
\def\eea{\end{eqnarray}}
\def\ba{\begin{array}}
\def\ea{\end{array}}

\graphicspath{{Figures/}}
\begin{document}

%:Title Info
\title{Quantum Information Approach to Dynamical Diffraction Theory}

% Joachim
\author{J. Nsofini} 
\email{jnsofini@uwaterloo.ca}
\affiliation{Department of Physics, University of Waterloo, Waterloo, ON, Canada, N2L3G1}
\affiliation{Institute for Quantum Computing, University of Waterloo,  Waterloo, ON, Canada, N2L3G1}
%Kamyar
\author{K. Ghofrani}
\affiliation{Institute for Quantum Computing, University of Waterloo,  Waterloo, ON, Canada, N2L3G1}  
%Dusan
\author{D. Sarenac}
\affiliation{Department of Physics, University of Waterloo, Waterloo, ON, Canada, N2L3G1}
\affiliation{Institute for Quantum Computing, University of Waterloo,  Waterloo, ON, Canada, N2L3G1} 
% David
\author{D. G. Cory}
\affiliation{Institute for Quantum Computing, University of Waterloo,  Waterloo, ON, Canada, N2L3G1} 
\affiliation{Department of Chemistry, University of Waterloo, Waterloo, ON, Canada, N2L3G1}
\affiliation{Perimeter Institute for Theoretical Physics, Waterloo, ON, Canada, N2L2Y5}
\affiliation{Canadian Institute for Advanced Research, Toronto, Ontario, Canada, M5G1Z8}
% Dmitry
\author{D. A. Pushin}
\affiliation{Department of Physics, University of Waterloo, Waterloo, ON, Canada, N2L3G1}
\affiliation{Institute for Quantum Computing, University of Waterloo,  Waterloo, ON, Canada, N2L3G1}

\begin{abstract}
We present a simplified model for dynamical diffraction of particles through a periodic thick perfect crystal based on repeated application of a coherent beam splitting unitary at coarse-grained lattice sites. By demanding translational invariance and a computationally tractable number of sites in the coarse-graining we show how this approach reproduces many results typical of dynamical diffraction theory and experiments. This approach has the benefit of being applicable in the thick, thin, and intermediate crystal regimes. The method is applied to a three-blade neutron interferometer to predict the output beam profiles, interference patterns, and contrast variation.
\end{abstract}


\maketitle

%==============================================================
% Introduction
%==============================================================
\section{Introduction }
\label{sec:intro}
Dynamical diffraction (DD) is a theory describing the interaction of photons and matter waves satisfying Bragg and near Bragg diffraction condition in perfect periodic crystal lattices~\cite{Zachariasen67, Batterman&Cole, Rauch&Petrascheck78,Sears78, Authier2006,Darwin1914,James1963,Reimer1984,Oberthaler1999}.  
It has been used to explain and predict many features of diffraction from periodic lattices; for example  Pendell{\"o}sung oscillations \cite{Abov,Shull63}, the extinction length and abnormal transmission \cite{Batterman&Cole} as well as the Borrmann effect \cite{bormann}. An incident wave satisfying the Bragg conditions is split by interactions with the atoms of a periodic crystal lattice. This is often approximated in the thick crystal regime by two waves emerging from the crystal, one propagating along the incident wave direction called the  transmitted or forward diffracted wave, and one propagating in the complementary direction called the reflected or Bragg diffracted wave. The properties of each of these waves are dependent on the momentum at incidence and the nature of the lattice \cite{Zeilinger81,Abov,Lemmel2013}.

While DD theory has been very successful for explaining many diffraction phenomena, the mathematics can be quite cumbersome and involve solving the Schr{\"o}dinger's equation for a lattice with Avogadro's number of interaction potentials. Even in the two-wave approximation, the standard theory of DD  still uses many variable substitutions to make the formulae readable \cite{Lemmel2007,Sam,Vladimir}.  This lack of readability may end up obscuring very simple concepts.

Presented here is a brief review of the standard theory of DD, and then an alternative and relatively simple treatment of DD using the language of quantum information (QI) theory. This approach models a periodic lattice as a network of beam-splitters. Furthermore, it treats DD as a coherent quantum effect that arises due to the interference of different paths taken by the wave as it passes through the lattice. 
While comparatively simple it is still able to accurately explain many DD effects. We will consider beam profiles for DD through a single thick crystal, and show how it predicts the widening of the neutron beam profile. This widening is bounded by the outer path in the transmitted beam and the outer path in the reflected beam thereby forming a triangular region. This triangular region is known as the Borrmann triangle. This QI model also predicts the sinusoidal variations known as Pendell{\"o}sung oscillations in the intensities of transmitted and reflected beams. These Pendell{\"o}sung oscillations result from energy transfer between the reflected and transmitted beams. Lastly, it is shown how this approach may be extended to a multi-blade interferometric device such as the three-blade neutron interferometer.

%==============================================================
% Dynamical Diffraction
%==============================================================
\section{Dynamical Diffraction}
\label{sec:dd}
\begin{figure}
\center
\includegraphics[scale=0.32]{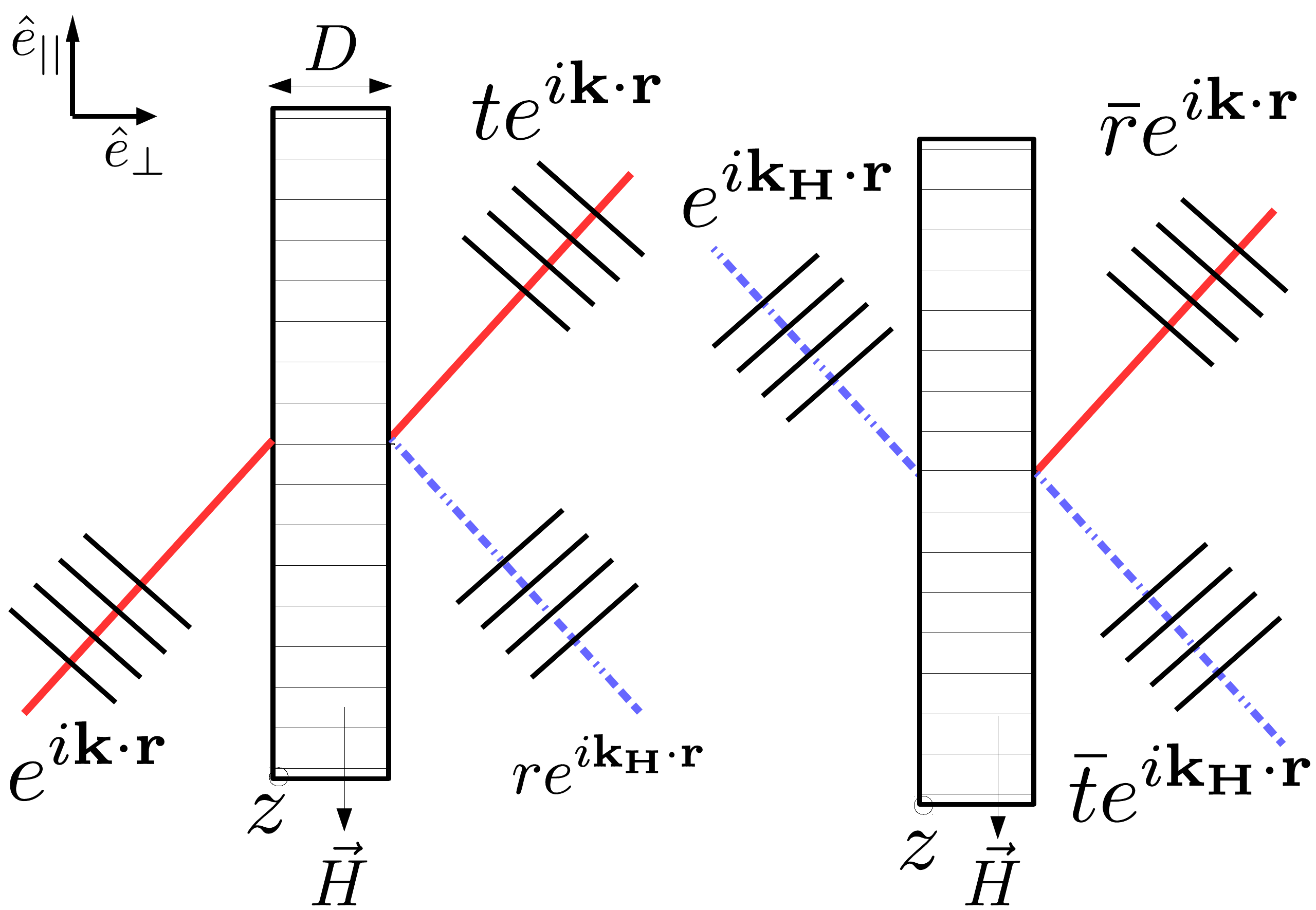}
\caption{Bragg diffraction from a crystal in the Laue geometry. In the Laue geometry the diffracted beam is on a different surface of the crystal relative to the incident beam. The two possible inputs to the crystal lead to different transmission and reflection coefficients for the waves exiting the crystal.} \label{Laue}
\end{figure}

For completeness a brief review of the standard theory of DD is presented. Consider a particle described by a wave-function $\Psi(\mathbf{r}) = \int d\mathbf{k} \,\mu_{\textbf{k}}\,\psi_{\mathbf{k}}(\mathbf{r})$, where $\psi_{\mathbf{k}}(\mathbf{r})=e^{i\mathbf{k\cdot r}}$ are a basis of plane-waves, and $\mu_{\mathbf{k}}$ describes the particle's momentum distribution. Let this particle be incident on a perfect periodic crystal of thickness $D$  located at position $\mathbf{r}=z\hat{e}_\perp$. The crystallographic orientation is assumed to be perpendicular to the crystal surface such that $\mathbf{k}={k}_\perp \hat{e}_\perp+k_\parallel \hat{e}_\parallel$, where, $\hat{e}_\perp (\hat{e}_\parallel)$ are unit vectors perpendicular (parallel) to the crystal surface (see Fig.~\ref{Laue}). In this configuration (commonly referred to as the Laue geometry) the transmitted beam and the reflected beam both exit from the same surface of the crystal. Inside the crystal the wavefunction $\Psi(\mathbf{r})$ must  satisfy the stationery state Schrodinger equation 
\bea
\label{SEq}
\left[-\frac{\hbar^2}{2m}\nabla^2+V(\mathbf{r})\right]\Psi(\mathbf{r})=E\Psi(\mathbf{r}),
\eea
where the potential $V(\mathbf{r})$ describes the scattering centres in the crystal. For a periodic crystal $V(\mathbf{r})=V(\mathbf{r+R})$, where $\mathbf{R}$ is the translation vector inside the crystal. The energy $E$ is the total energy of the particle inside the crystal and is equal to the kinetic energy of the particle in free space, $E_0=\hbar^2k_0^2/{2m}$, where $m$ and $k_0$ are the particle mass and the free space wavevector associated with the particle. The solution to Eq.~(\ref{SEq}) are multiple scattering waves. Due to the vanishingly small nature of the interaction potential ($V\ll E$), two energies $E_K$ and $E_{K_H}$ are excited from the periodic lattice at the Bragg condition. The corresponding eigenstates of Eq.~(\ref{SEq}) are double degenerate Bloch waves. As a result, four waves propagate inside the crystal with two wave vectors in the transmitted and two in the reflected directions. The beating of these waves generates a fine feature of dynamical diffraction, namely the Pendell{\"o}sung oscillation. Using the boundary conditions  the four waves recombine to two waves as they exit the crystal giving rise to the transmitted $\psi_O=t\psi_k$ and the reflected $\psi_H=r\psi_{k_H}$ waves. In this geometry, O and H generally refer to the the forward diffracted (transmitted) and Bragg diffracted (reflected) directions respectively (as defined by the reciprocal lattice vector $\vec{H}$). 
The reflection ($r$) and transmission ($t$) coefficients for non absorbing crystals  are given by~\cite{Sam,Lemmel2013}
\bea
\label{ddeq}
t 	&=& 	e^{i\chi}\exp\left(-iA\eta\right)\bigg\lbrace\cos(A\sqrt{1+\eta^2}) \nonumber\\ 
	& &	+\frac{i\eta}{\sqrt{1+\eta^2}}\sin(A\sqrt{1+\eta^2})\bigg\rbrace,\\
r	&=&	e^{i\chi}\exp\left[i(-A\eta+2A\eta z/D)\right]   \left(\frac{v_H}{v_{-H}}\right) \label{ddeq2}\nonumber\\ 
	& &	\times \frac{-i}{\sqrt{1+\eta^2}}\sin(A\sqrt{1+\eta^2})
\eea
where the phase shift $\chi=D(K_\perp-k_\perp)$, $K_\perp = \sqrt{k_\perp^2-2mV/\hbar^2}$,  is the nuclear phase shift due to the crystal (note that this phase shift also occurs outside the Bragg conditions), $A=\pi D/\Delta_{H}$ is the dimensionless crystal thickness, $\Delta_{H}=K_\perp\hbar^2\pi |V_H|^{-1}m^{-1}$ is the extinction length, and $\eta=\frac{1}{2}(E_{K_H}-E_K)/|v_H|$ parametrizes the normalized energy difference of $E_{K_H}$ and $E_K$. $v_H$ is the Fourier component of the potential. In general, $\eta$ characterizes the deviation from the exact Bragg condition and $\eta$ is commonly known as beam divergence or momentum (wavelength) spread.

%:	Figure: Bragg Diffraction

Using the transmission and reflection amplitudes in Eq.~(\ref{ddeq},\ref{ddeq2}) we define an effective unitary operator for a  blade of silicon [220] written in matrix form as  
\bea
U_{DD}({\varphi,\varrho,\vartheta})&=&e^{i\chi}\begin{pmatrix}
 e^{i\varphi}\cos\vartheta & e^{i\varrho}\sin\vartheta\\
-e^{-i\varrho}\sin\vartheta & e^{-i\varphi}\cos\vartheta
 \end{pmatrix}
 \label{eq:DDUnitary}
\eea
where, from Eq.~(\ref{ddeq}) we get
\begin{align}
\varphi &\equiv -A\eta+\arctan\left[\frac{\eta}{1+\eta^2}\tan[\Phi(\eta)]\right],
\label{Eqn:DDtoQI1}\\
\varrho &\equiv -A\eta+2A\eta z/D+\pi/2,
\label{Eqn:DDtoQI2}\\
\vartheta &\equiv \arcsin\left[ \frac{\sin[\Phi(\eta)]}{\sqrt{1+\eta^2}}\right],
\label{Eqn:DDtoQI3}
\end{align}
with $\Phi(\eta)=A\sqrt{1+\eta^2}$. Under ideal conditions in DD (no momentum spread) the parameters are  $\varphi=0,\varrho=\pi/2,\vartheta=A$ leading to an overall unitary of the blade that is independent of $\varphi, \varrho$ and $\vartheta$. In this case the neutron wavefunction at the exit of a crystal carries no phase information about the crystal.

In sec.~\ref{sec:ddqi} we extend the definition of the unitary operator to develop a model for dynamical diffraction based on quantum information.

%==============================================================
% QI Model of DD
%==============================================================
\section{Quantum Information Model for Dynamical Diffraction} \label{sec:ddqi}

The process of DD  through a perfect periodic non-absorbing crystals is a unitary process. In this work a proposed alternative quantum information model for DD based on the requirements that a crystal can be segmented  into  planes each acting as a unitary operator, and the same unitary operator is repeatedly applied through out the process. This is an operational approach that considers a coarse-graining of a thick perfect crystal into a computationally tractable number of planes of logical scattering sites. Each of the logical scattering sites may be modelled as a general beam splitter that coherently splits an incoming wave into transmitted and reflected components. The choice of unitary depends on the number of planes in the coarse-graining with the choice made so that the phase of the wavefunction leaving the crystal does not wrap around in mutlple of 2$\pi$'s. As the number of planes increases, so does the number of possible paths through the crystal. This results in a widening of the matter wave beam profile, and the interference between these multiple paths reproduces many of the effects typically described by standard DD theory. The coarse-graining of scattering sites is necessary since there are an order of Avogadro's number of atoms corresponding to physical scattering sites in a perfect crystal. However, many DD effects can be reproduced with only a modest number of coarse-grained scattering sites considered. This QI approach is a quantum version of a Galton's board, which is one form of a discrete time quantum walk \cite{Galton,qw,QWalk}.  The model is also related to the original proposal of DD by Darwin \cite{Darwin1914, Darwin1922} that involves breaking down the scattering media into layers in conformation to the invariance principle \cite{Abov, Ambartsumyan81}. 

In the QI model, it is assumed that the scattering at each coarse-grained site is macroscopically distinct, similar to Bragg scattering and Bloch theory. This model also makes the assumption that the entire process of diffraction through the crystal is within the coherence length of the incoming wave.

\begin{figure}
\center
\includegraphics[scale=.35]{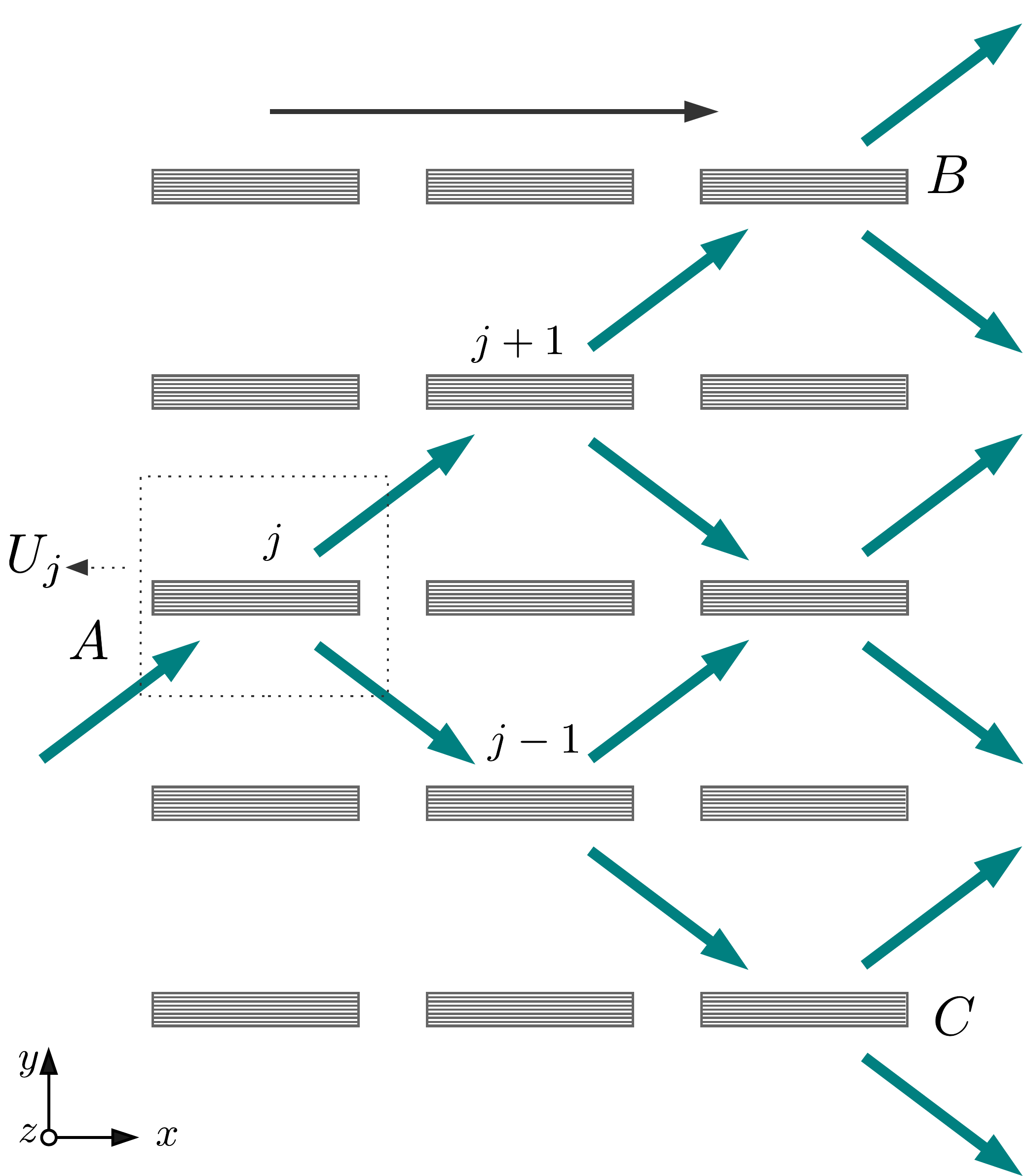}
\caption{A perfect crystal decomposed into various scattering sites (nodes) shown as blocks. The incident wave is a single ray moving upward along $AB$. The direction $AC$ is the reflected wave. The region $ABC$ is the Borrmann triangle. At each node the unitary operator $U_j$ coherently splits the wave into two components; the transmitted and reflected. Repeated application of $U_j$ generates the transmitted and reflected outputs of the crystal. Any ray of the transmitted beam has undergone an even number of reflections while that of the reflected beam has undergone an odd number of reflections.}
 \label{fig:FBlade}
\end{figure}

The coarse-graining for a perfect crystal into scattering nodes is illustrated in Fig.~\ref{fig:FBlade}. This coarse-graining procedure is performed in several steps. The crystal is segmented into planes, and each plane is further divided into \emph{nodes} creating a lattice where each node corresponds to a scattering site. Each node functions as a beam-splitter with two input and two output ports. Hence the two incident paths scatter to the \emph{transmitted} and \emph{reflected} components. The scattering action of an arbitrary node (denoted by $j$ which specifies the node's relative vertical location) may therefore be modelled as a unitary transformation $U_j$ acting on two levels, $|a_j\rangle$, $|b_j\rangle$ of the incident state. The collective action of all the nodes leads to a multiple scattering process that gives rise to quantum interference effects. The unitary generating this multi-scattering process is denoted by ${\cal U}(N)$, and its dimensions depend on the number of planes, $N$, considered in the coarse grained approximation. The properties of the transmitted and reflected waves leaving the crystal surface, as well as the corresponding intensities, depend on the nature of ${\cal U}(N)$.

A ray or single line is used to represent a single logical level of any wave, which may be modelled as a state vector $|{a_j}\rangle$ or $|{b_j}\rangle$, where the labels $a$ and $b$  refer to rays moving upwards (positive $y$-momentum) or downwards (negative $y$-momentum) respectively. The ray tracing approach is analogous to a path integral and is used to illustrate some of the features of the wave leaving the crystal.

%:	Figure: Blade

At the scattering site the beams are coherently split according to the relation
\begin{align}
\ket{a_j} &\mapsto t_{a} \ket{a_{j+1}} + r_a \ket{b_{j-1}}	\\
\ket{b_j} &\mapsto r_{b} \ket{a_{j+1}} + t_b \ket{b_{j-1}}
\end{align}
where $t_a, t_b, r_a, r_b$ are complex transmission and reflection coefficients. For the scattering relation to be unitary the following conditions have to be satisfied:
\begin{align}\nonumber
1 = |t_a|^2 + |r_a|^2,\quad
1 = |t_b|^2 + |r_b|^2,\quad
0 = t_a \overline{r_b} + r_a \overline{t_b}.
\end{align}
A possible choice for the coefficients is 
\begin{align}
t_a &= e^{i \xi} \cos\theta   	&t_b &= e^{-i \xi} \cos\theta	\nonumber\\
r_a &= -e^{-i \zeta}\sin\theta 	&r_b &= e^{i \zeta}\sin\theta.
\label{eqn:tcoeff}
\end{align}

A range of meaningful values for these parameters may be gotten from the minimum thickness of a crystal, $\tau$, for which the particle wavefunction reflected. According to the standard DD theory this minimum thickness is related to the Pendell{\"o}sung length, $\Delta_H$, by the ratio, $\tau/\Delta_H<1/2$. By comparing this ratio to the parameters in the QI model, we obtain that $\theta=\pi\tau/\Delta_H$, $\xi=0,$ and $\zeta=\pi$. For a specific situation $\theta$ can be optimized to reproduce experimental results. 

The unitary operator is,
\begin{align*}
U_{j,\xi,\theta,\zeta}
	&= \left(e^{i \xi}\cos\theta\ket{a_{j+1}}-  e^{-i \zeta}\sin\theta\ket{b_{j-1}}\right)\bra{a_j} \nonumber\\
	&+\left(e^{i \zeta}\sin\theta\ket{a_{j+1}} +  e^{-i \xi}\cos\theta\ket{b_{j-1}}\right)\bra{b_j} \nonumber\\
	&=  \ket{a_{j+1}}\left(e^{i \xi}\cos\theta\bra{a_j}+  e^{i \zeta}\sin\theta \bra{b_j}\right)\nonumber\\
	&-\ket{b_{j-1}}\left(e^{-i \zeta}\sin\theta\bra{a_j}-  e^{-i \xi}\cos\theta\bra{b_j}\right),
\end{align*}
which will be represented as $U_j$ from now on.
%Or expressed as a matrix
%\bea
%U_{\xi,\theta,\zeta}&=&\begin{pmatrix}
% e^{i\xi}\cos\theta & e^{i\zeta}\sin\theta\\
%-e^{-i\zeta}\sin\theta & e^{-i\xi}\cos\theta
% \end{pmatrix},
%\eea
%with the subscript $j$ dropped because the unitary is independent of the location.
In this parametrization $\xi$ and $\zeta$ are the phases of the transmitted and reflected beams respectively, and $\theta$ determines the relative probability of the reflected and transmitted beams from a single node. 

Consider a normalized input beam $\ket{\Psi_0} = \alpha \ket{\Psi_0^T} + \beta \ket{\Psi_0^R}$ spanning multiple nodes, where the upward propagating $\ket{\Psi_{0}^{T}}$ and the downward propagating $\ket{\Psi_{0}^{R}}$ components are given by
\begin{align}
\ket{\Psi_{0}^{T}}&=\sum_{j}\alpha_j \ket{a_j}, 
\quad \ket{\Psi_{0}^{R}}= \sum_{j} \beta_j \ket{b_j}.
\label{eq:beamcoeff}
\end{align}
From the normalization condition $|\alpha|^2 + |\beta|^2 =1$,
\[
 \sum_{j}|\alpha_j|^2 + \sum_{j}|\beta_j|^2 =1.
\]
Then the action of all the scattering nodes of a single vertical plane is given by
\begin{align}
U_j\ket{\Psi_{0}^{T}} &=  \sum_{j\in T}\alpha_j\Big(t_a \ket{a_{j+1}}+ r_a \ket{b_{j-1}}\Big)	\\ 
U_j\ket{\Psi_{0}^{R}} &= \sum_{j\in R}\beta_j \Big(r_b \ket{a_{j+1}} - t_b \ket{b_{j-1}}\Big),
\end{align}
so that the (unnormalized) transmitted and reflected beams are given by
\begin{align}
\ket{\Psi_{1}^{T}} &= t_a\left( \sum_{j\in T}\alpha_j \ket{a_{j+1}}\right)
				 +r_b\left( \sum_{j\in R}\beta_j \ket{a_{j+1}}\right), \label{Eq:TWFblade}
\\
\ket{\Psi_{1}^{R}} &= r_a\left( \sum_{j\in T}\alpha_j \ket{b_{j-1}}\right)   
				-t_b\left(\sum_{j\in R}\beta_j  \ket{b_{j-1}}\right). 
\label{Eq:RWFblade}
\end{align}
These states are then the input to the next plane. The process is repeated until the last plane at the exit surface is reached. The appropriate normalization factor for the transmitted and reflected beams will depend on the reflection and transmission coefficients and the resulting interference.

In general,  after propagating though a crystal segmented in to $N$ planes the components of the wavefunction are
\begin{align}
%\ket{\Psi_{N}}&={\cal U}\ket{\Psi_0},
\ket{\Psi_{N}^{T}}&=\sum_{j\in T}\alpha_j \ket{a_j},\quad
 \ket{\Psi_{N}^{R}}= \sum_{j \in R} \beta_j \ket{b_j},
\label{eq:beamcoefAfterN}
\end{align}
where $\alpha_j=\bra{a_j}{\cal U}(N)\ket{\Psi_0}, \beta_j=\bra{b_j}{\cal U}(N)\ket{\Psi_0}$. Using these equations the effective unitary of a single blade ${\cal U}(N)$ based on the QI model can be derived.

The QI model can be used to derive results consistent with DD, examples included are
\begin{itemize}
\item Wave field in the Borrmann triangle.
\item Integrated intensities after diffraction.
\item Pendell{\"o}sung oscillations.
\end{itemize}
These are described in Section~\ref{sec:dd-app}. In addition, this approach provides a simple phenomenological way to study various types of noise processes considered in quantum information theory, such as dephasing, that may occur during a diffraction process due to variations in the parameters of individual scattering nodes when averaged across many particles. It is also simple to generalize this approach to predict the behaviour of multi-blade devices such as a neutron interferometer, which is discussed in Section~\ref{sec:dd-ni}.

%==============================================================
% Beam Splitter
%==============================================================
\subsection{Example: 50:50 Beam Splitter}
Here we apply the presented formalism to the particular case where each node acts as a 50:50 beam splitter. Considered the parameters $\xi=0,\theta=\pi/2,\zeta=0$ which set the unitary $U_j$ equal to the Hadamard matrix
\[
U_j=\frac{1}{\sqrt{2}}  \ket{a_{j+1}}\big(\bra{a_j}+ \bra{b_j}\Big)
	+\frac{1}{\sqrt{2}} \ket{b_{j-1}}\Big(\bra{a_j}-\bra{b_j}\Big).
\]
%
%\[
%U_j=\frac{1}{\sqrt{2}}\begin{pmatrix}
% 1 & 1\\
%1 & -1
% \end{pmatrix}.
% \]
This is equivalent to nodes acting as 50:50 beam splitters. For an input state $\ket{\Psi_0}= \ket{\psi_j} = \alpha_0 \ket{a_j} + \beta_0 \ket{b_j}$, the output from the single node is given by
\[
\ket{\Psi_1}=U_j\ket{\psi_j} = \left(\frac{\alpha_0 +\beta_0}{\sqrt2}\right)\ket{a_{j+1}} + \left(\frac{\alpha_0 -\beta_0}{\sqrt2}\right)\ket{b_{j-1}}.
\]
 
Consider the case of a single input ray in the state $\ket{\psi_j} =\ket{a_j}$ onto a scattering node $j$. After the node the state is 

\[
\ket{\Psi_1}=\frac{1}{\sqrt2}\left(\ket{a_{j+1}} + \ket{b_{j-1}}\right).
\]
At the second vertical plane the transmitted state $\ket{a_{j+1}}$ becomes an input to a node with unitary  $U_{j+1}$ and the reflected state $\ket{b_{j-1}}$ an input to a node with unitary  $U_{j-1}$:
\begin{align}
U_{j+1}\ket{a_{j+1}} 	&= \frac{1}{\sqrt2}\Big(\ket{a_{j+2}} +\ket{b_{j}}\Big),	\\
U_{j-1}\ket{b_{j-1}} 	&= \frac{1}{\sqrt2}\Big(\ket{a_{j}} -\ket{b_{j-2}}\Big),
\end{align}
so that 
\begin{align}
\ket{\Psi_2} 
		&= U_{j+1}U_{j-1}\ket{\Psi_1} = U_{j+1}U_{j-1}U_j\ket{\Psi_0},\\
		&= \frac{1}{\sqrt{2}}\Big(\ket{\Psi_2^T}+\ket{\Psi_2^R}\Big),
\end{align}
where the transmitted beam $\ket{\Psi_2^T}$ and reflected $\ket{\Psi_2^R}$ beam each consists of two rays
\begin{align}
\ket{\Psi_2^T} &=  \frac1{\sqrt2}\Big(\ket{a_{j+2}} +\ket{a_j}\Big) \\
\ket{\Psi_2^R} &=  \frac1{\sqrt2}\Big(\ket{b_{j}} -\ket{b_{j-2}}\Big).
\end{align}

Adding an additional plane to make the three-plane case results in three rays in each of the transmitted and reflected beams 
\begin{align}
\ket{\Psi_3} 
		&= \sqrt{\frac23}\ket{\Psi_3^T} + \frac1{\sqrt3}\ket{\Psi_3^R}	\\
\ket{\Psi_3^T}	&= \frac{1}{\sqrt6}\Big( \ket{a_{j+3}} +2\ket{a_{j+1}} - \ket{a_{j-1}} \Big)	\\
\ket{\Psi_3^R}	&= \frac{1}{\sqrt{2}}\Big( \ket{b_{j+1}}+ \ket{b_{j-3}} \Big).
\end{align}
Due to constructive and destructive interference of the state $\ket{a_{j+1}}$ and $\ket{b_{j-1}}$, two-third of the intensity is in the transmitted beam, and one-third in the reflected beam.

Adding a fourth plane gives 
\begin{align}
\ket{\Psi_4} 
		&= \sqrt{\frac34}\ket{\Psi_4^T} +\frac12\ket{\Psi_4^R}
\label{eq:beamcoefAfter4}		
			\\
\ket{\Psi_4^T}	&= \frac{1}{2\sqrt3}\Big( \ket{a_{j+4}} +3\ket{a_{j+2}} - \ket{a_{j}} + \ket{a_{j-1}} \Big)
\label{eq:TbeamcoefAfter4}
	\\
\ket{\Psi_4^R}	&= \frac{1}{2}\Big( \ket{b_{j+2}} +\ket{b_{j}} - \ket{b_{j-2}} - \ket{b_{j-4}} \Big).
\label{eq:RbeamcoefAfter4}
\end{align}
Just  with four planes and a 50:50 beam splitter we notice that the beam at the exit spreads unevenly due to interference.

In general,  after propagating though a media with $N$ vertical planes the transmitted and reflected components of the  wavefunction are
\begin{align}
%\ket{\Psi_{N}}&={\cal U}\ket{\Psi_0},
\ket{\Psi_{N}^{T}}&=\sum_{j\in T}\alpha_j \ket{a_j},\quad
 \ket{\Psi_{N}^{R}}= \sum_{j \in R} \beta_j \ket{b_j},
\label{eq:beamcoefAfterN}
\end{align}
where in general the probability amplitudes $\alpha_j\neq\beta_j$.  

It is possible to use this QI model to extract information about parameters in DD experiments. This is captured by Eqs.~(\ref{Eqn:DDtoQI1},\ref{Eqn:DDtoQI2},\ref{Eqn:DDtoQI3}) with functional dependence of $\xi,\zeta,\theta$.  Various applications of DD and choices of these parameters are considered in the next section.

%:	Figure: Intensity Profiles
\begin{figure}
\center
\includegraphics[width=\columnwidth]{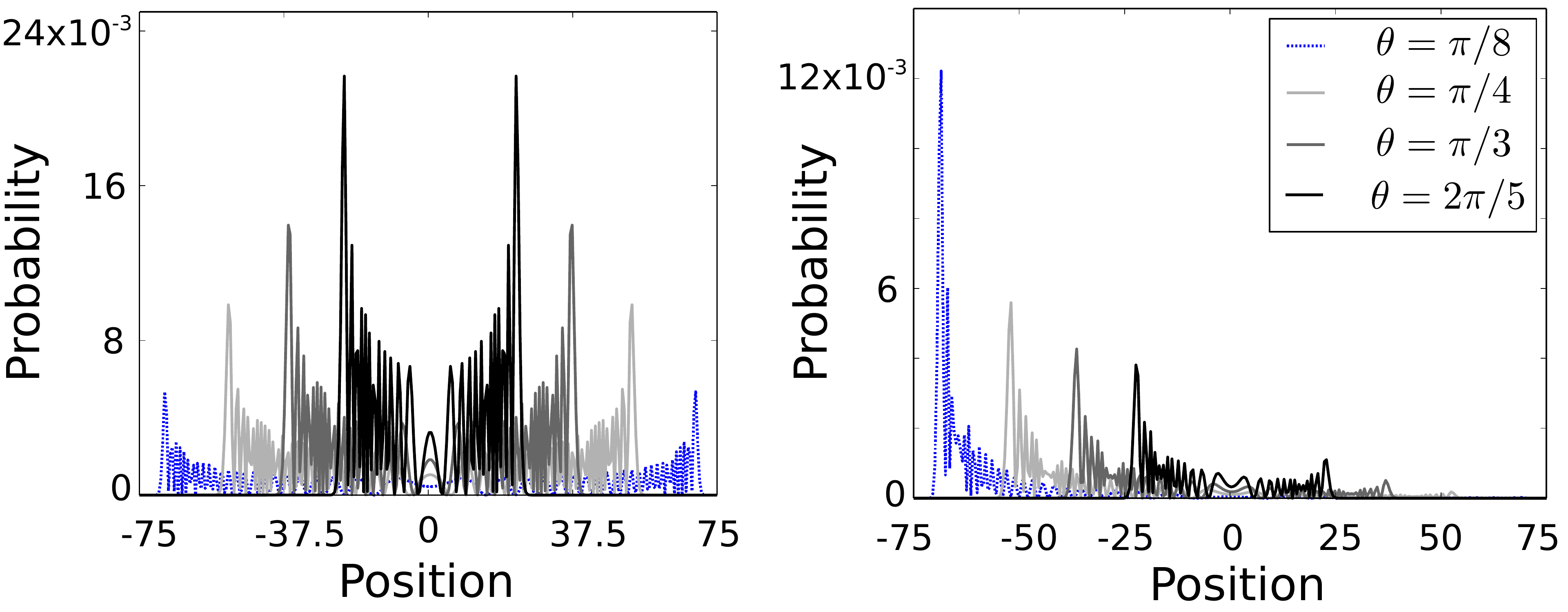}
\caption{Intensity profiles for the reflected (left) and transmitted (right) beams for a thick crystal modelled by $N=150$  scattering planes, and for  transmission and reflection coefficients $|t|=\cos\theta, |r|=\sin\theta$, for values of $\theta=\pi/8,\pi/4,$ $\pi/3, 2\pi/5$. The reflected beam (left) is symmetric with two peaks at the edges, while the transmitted (right) beam is asymmetric with a single peak at the outside. As $|t|$ approaches 1 the widths of both beams are compressed.
} \label{fig:FBman}
\end{figure}
%==============================================================
% Applications
%==============================================================
\section{Applications}
\label{sec:dd-app}
In this section the QI model is applied to some well known phenomenon in DD. Through out this section, the state of the neutron at the input (node at $j=0$) is
\[ \ket{\Psi_0}=  \ket{a_0},
\]
which is a single ray propagating upwards.

%-------------------------------------------------------------------------------------------------------------
% Borrman Triangle
%-------------------------------------------------------------------------------------------------------------
\subsection{Intensity Profile of the Borrmann Triangle}

The first application considered is a simulation of the position dependent intensity profile for a single crystal. The spreading of these profiles caused  by the crystal thickness is known as the \emph{Borrmann fan}, and it has been observed experimentally by scanning a slit of several microns wide across the output surface of the crystal~\cite{Shull63}. The triangle formed by the outer edges of the transmitted beam, reflected beam, and the input point of a single ray is called the \emph{Borrmann triangle}. In QI model it is given by the region $ABC$ in Fig.~\ref{fig:FBlade}. ${AB}$ is along the transmitted wave direction  while $AC$ is along the reflected wave direction.
As expected the intensity profile of the transmitted and reflected beams exiting the crystal depends on both the number of planes considered in the model, and the transmission and reflection coefficients $t_a, t_b, r_a, r_b$ for a single node. 

The intensity spreading in the Borrmann triangle has been observed experimentally \cite{Shull73a,Shull63}. To study this in the QI model the intensities at the output of the crystal are simulated for various values of $\theta$. The transmitted and reflected probability at the output node $j$ is given by
\bea\nonumber
I^T_j=|\bra{a_j}{\cal U}(N)\ket{a_0}|^2,\\ I^R_j=|\bra{b_j}{\cal U}(N)\ket{a_0}|^2.
\eea
Figure~\ref{fig:FBman}  shows the reflected and transmitted intensity distributions of the exiting beam across the crystal surface for $N=150$ and various values of $\theta =\pi/8$, $\pi/4$, $\pi/3$, $2\pi/5$. This figure illustrates that the reflected beam has a symmetric profile with two intensity peaks at the edges of the beam while the transmitted beam is asymmetric with the beam concentrated on one side. In addition, as the transmission coefficient for a single node approaches 1 (equivalently $\theta\rightarrow\pi$) the beam is compressed in width and the intensity in this region increases. The data presented in Figure~\ref{fig:FBman}, which is obtained with only a modest number of planes, is in good agreement with those obtained by the standard theory of DD and those observed experimentally.

%-------------------------------------------------------------------------------------------------------------
% Pendellosung Oscillations
%-------------------------------------------------------------------------------------------------------------
\subsection{Integrated Intensities and Pendell{\"o}sung Oscillations}
%transmitted and reflected  intensities converges to 1/2.
%==============================================================
%
%:	Figures: Integrated Int
\begin{figure}
\center
\includegraphics[width=\columnwidth]{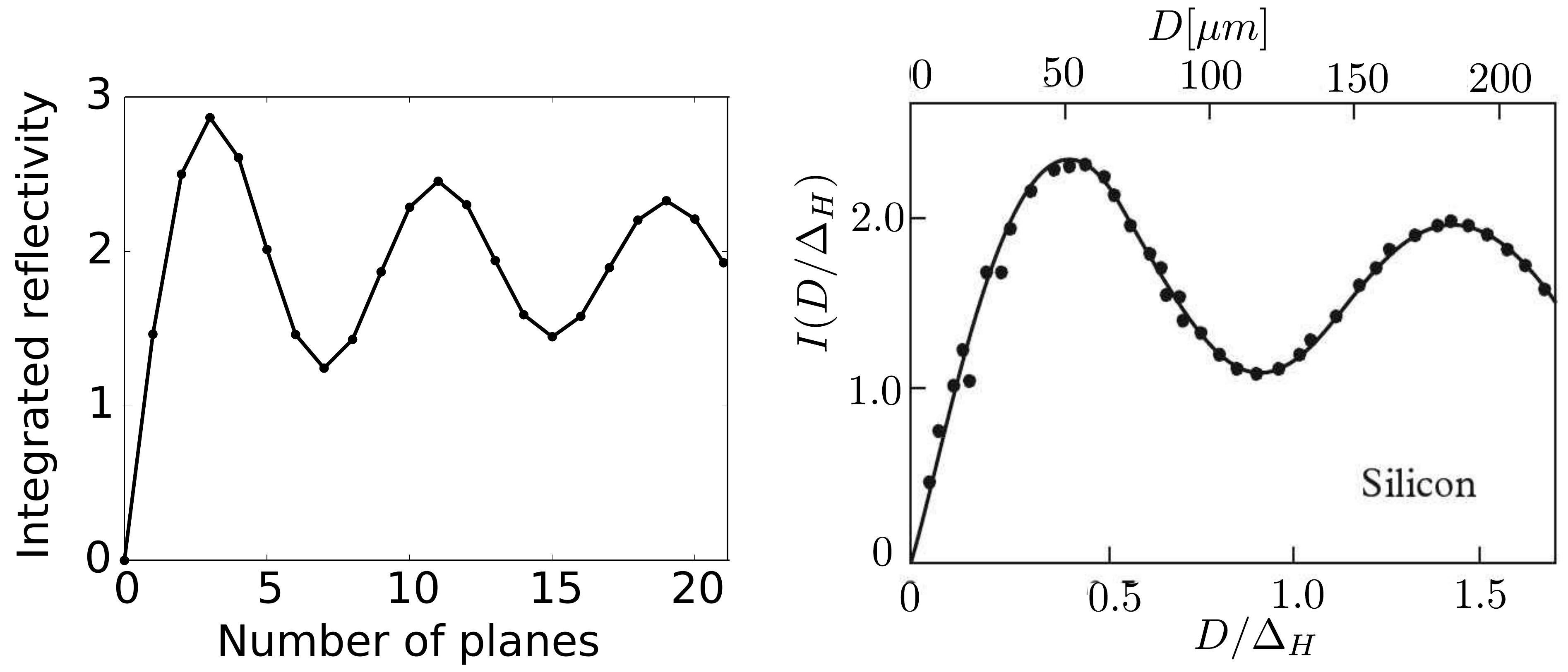}
\caption{The left figure is the integrated reflectivity at the exit of a single crystal based on the QI model with a node unitary operator $U_{j,0,0,\theta=\pi/8}$.
The right figure is the measured integrated reflectivity of Bragg scattering in the Laue diffraction, see Fig. 1 of ref. \cite{Sippel1965}. Copyright permission from Elsevier. In comparison a Si thickness of 50 $\mu$m corresponds to three planes in the QI model.} \label{SumIntRT}
\end{figure}
%==============================================================
%
In the second application, the quantum information model is used to reproduce the \emph{integrated intensities} of the output beam. The integrated intensities  are taken by summing all of the probabilities of each of the transmitted and reflected beams in the triangle. For crystal segmented into $N$ planes the relative  integrated transmitted intensity to the relative integrated reflected intensity allows us to define the integrated transmission $I_T$ and reflection $I_R$ coefficients as
\bea\label{Sumprob1}
I_T&=\sum_{j} |\bra{a_j}{\cal U}(N)\ket{a_0}|^2,\\ I_R&=\sum_{j}|\bra{b_j}{\cal U}(N)\ket{a_0}|^2.\label{Sumprob2}
\eea
The integrated transmission and reflection coefficients are known to undergo oscillations, called Pendell{\"o}sung oscillations. The integrated reflectivity was repeatedly observed in experiments where either the crystal thickness was varied \cite{Sippel1965,Somenkov1978}, or the neutron energy was varied~\cite{Shull1963}.

The integrated reflectivity from a perfect single crystal silicon can be measured  by varying its thickness while keeping other conditions (wavelength, crystallographic orientation, etc) unchanged. The integrated transmitted and reflected intensities are periodic functions which are out of phase with each other. The  phase difference arises because the reflected and transmitted beams undergo an even and odd number of reflections respectively.

To simulate this variation in the QI model the parameters ${\xi,\theta,\zeta}$ are kept fixed  while the number of planes $N$ is varied. Figure~\ref{SumIntRT} shows side by side the integrated intensity predicted by the QI model  with a node unitary operator $U_{j,0,0,\pi/8}$ and the experimentally observed intensity~\cite{Sippel1965, Somenkov1978}. The two figures show a significant agreement leading to a plausible conclusion that a thickness of 50 $\mu$m corresponds to three planes for the specific wavelength and Si crystal used in the experiment.

Note that in the special case of a 50:50 splitting at the nodes with $\theta=\pi/4$, the normalized integrated intensities converges to 0.65 and 0.35 for the transmitted and reflected beams respectively, as the number of planes increases.

%\subsection{Pendellosung oscillations}
 In the theory of DD the probability current inside the crystal propagates in two components. One centred on the atomic planes position and the other at the inter-planar position. As the wave propagates through the crystal, these currents constantly exchange energy with one another in such a way that the total current is conserved. The energy exchange happens in an oscillating manner known as Pendell{\"o}sung oscillation~\cite{Sam}. In the theory of dynamical diffraction, Pendell{\"o}sung oscillations are best represented by plotting the reflected or transmitted intensity as a function of the deviation from the Bragg condition (the parameter $\eta$ mentioned in Sec.~\ref{sec:dd}), see \cite{Sudo06} for more details.  In the work of Shull \cite{Shull63}, the neutron energy was varied for three silicon single crystals of different thickness' and position sensitive detection implemented with a cadmium slit.

\begin{figure}
\center
\includegraphics[width=\columnwidth]{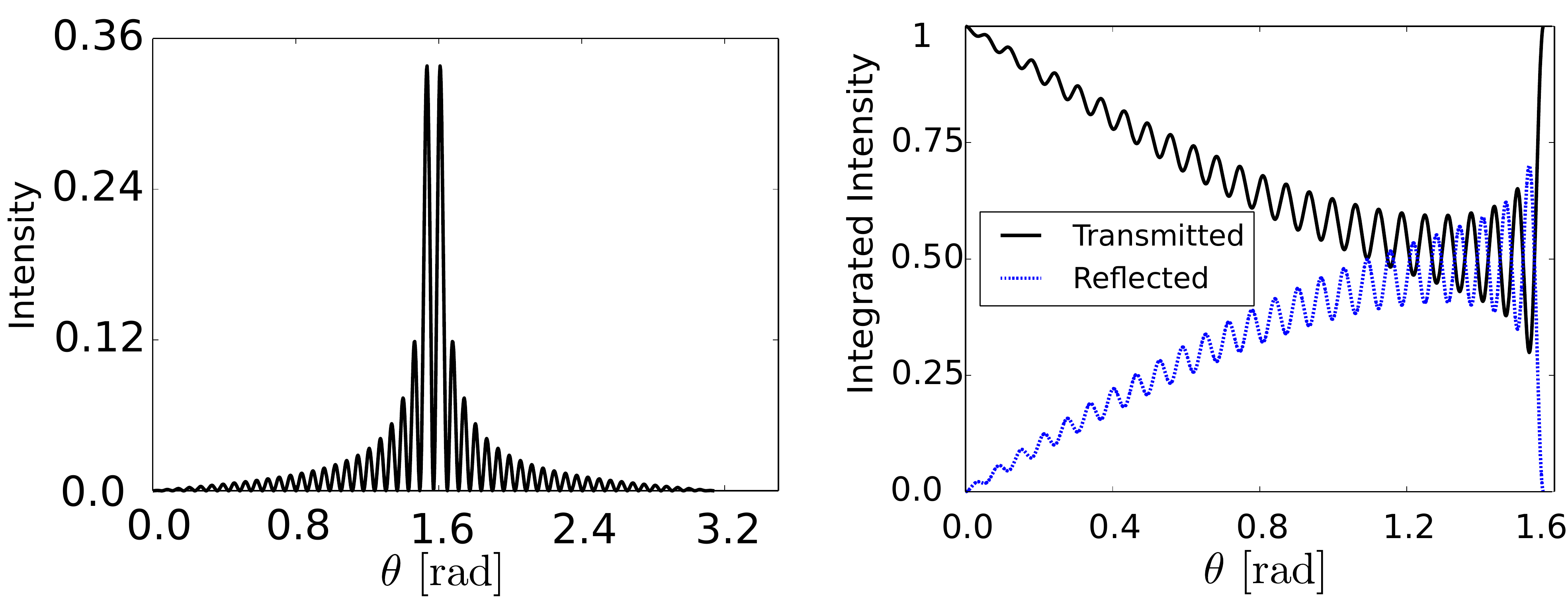}
\caption{Left figure is the simulated Pendell{\"o}sung oscillations at the exit of the Borrmann fan for the reflected beam. The QI model was done for a crystal with $N=50$ planes, and the node $j=25$ was post selected on. The intensity is in agreement with the conventional dynamical diffraction theory. The right figure shows the integrated intensities at the output as a function of $\theta$. As $\theta\rightarrow\pi/2$, both the reflected and transmitted intensities approach $1/2$. } \label{PenOsc}
\end{figure}

In order to simulate the Pendell{\"o}sung oscillations with the QI model the output intensity is post-selected on a specific node and the angle $\theta$ is varied to mimic the energy variation; while the number of planes $N$ is fixed. This is illustrated on the left plot of Fig.~\ref{PenOsc} for $N=50$, the unitary $U_{j,0,0,\theta}$ and $\theta\in [0,\pi]$. In this simulation the node $j=25$ was post selected on. These plots are consistent with the plots of the Pendell{\"o}sung oscillations obtained by the standard DD theory. On the right plot of Fig.~\ref{PenOsc} are the integrated intensities Eqs.~(\ref{Sumprob1},\ref{Sumprob2}) as a function of $\theta$. From the plot it can be noted that as $\theta\rightarrow\pi/2$ the integrated reflected and transmitted intensities both approach $1/2$.

%==============================================================
%
%:	Figures: Pendellosung Oscillation
%==============================================================
%
%==============================================================
% Extension to NI
%==============================================================
\section{ Extension to a  Neutron Interferometer}
\label{sec:dd-ni}
%%%%===============================================================================================================
%:	Figures: NI profile
\begin{figure}
\center
\includegraphics[scale=0.5]{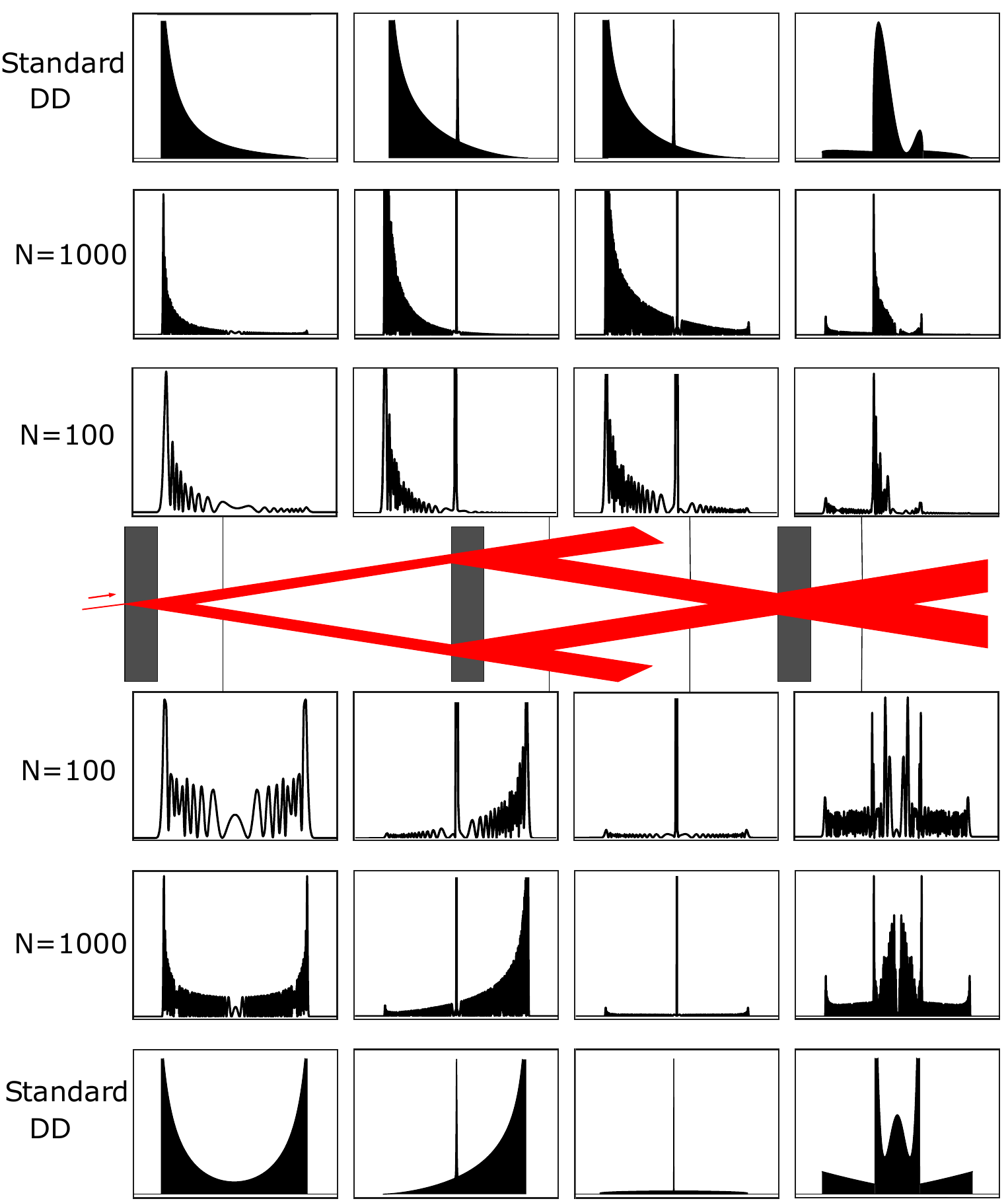}
\caption{The various intensity profiles for a three blade neutron interferometer. The inteferometer geometry and the beam trajectories are shown in the middle. The two inner rows correspond to our simulated profiles with $N=100$. The next two outer rows correspond to to our simulated profiles with $N=1000$. And for comparison, in the two outermost rows, the intensity profiles obtained by standard theory of dynamical diffraction. } \label{Ubonse}
\end{figure}
%%%%===============================================================================================================

The QI model of a single perfect crystal blade is extended to a three-blade and four-blade perfect crystal neutron interferometer (NI). The four-blade has the demonstrated advantage that is refocuses noise originating from mechanical vibration \cite{Dima2011}. For a concise application of the standard theory of DD to a neutron interferometer see \cite{Ubonse,Sam}. The NI is made of three blades of equal thickness which are machined from an ingot of single crystal silicon so that the Bragg planes of the blades are aligned. The first blade of the NI is identical to our single blade treatment and splits the neutron into the reflected and transmitted beams. The second NI blade splits the reflected and transmitted beams of the first blade into two other transmitted and reflected beams so that total of four beams emerge after the second blade. However only the transmission-reflection and reflection-reflection beams remain in the interferometer. Finally, these two beam paths are coherently recombined at the third blade. The third blade acts as an analyser as it coherently recombines the two beams allowing the device to function as a Mach-Zehnder interferometer. The two beams exiting the interferometer are historically labelled as \textit{O-beam} (which propagates along the same direction as the input beam) and \textit{H-beam}.

With the unitary operator of the blade defined from the standard theory of DD Eq.~(\ref{eq:DDUnitary}) and a non-dispersive phase difference $\chi$ between the two paths inside the interferometer, the wavefunction at the output of the NI is
\begin{align}
\ket{\Psi}&=\ket{\Psi^O}+\ket{\Psi^H},
\end{align}
where the O- and H- beam wavefunctions are
\begin{align}
\label{Eq:3bladeO}
\ket{\Psi^O}&=-e^{-i\varphi/2}\cos\vartheta\sin^2\vartheta\left(e^{-i\chi/2}+e^{i\chi/2}\right)\ket{a}\\
\ket{\Psi^H}&=ie^{i\varphi/2}\left(\cos^2\vartheta\sin\vartheta e^{i\chi/2}-\sin^3\vartheta e^{-i\chi/2}\right)\ket{b}.
\label{Eq:3bladeH}
\end{align}
When a neutron propagates through the NI with $\chi=0$, the beam  exits the NI with different probabilities that depends on $\vartheta$. For a balanced beam splitter $\vartheta=\pi/2$ the neutron emerges only through the O-beam.
Now we will use the QI model to simulate the beam profiles and the contrast for a three blade NI.

\subsection{{Beam Profiles}}
The neutron beam profiles produced by the QI model for each of the eight beams in the three blade NI are presented in Fig.~\ref{Ubonse}. For the simulation the unitary operator at each node is $U_{j,0,0,\theta=\pi/4}$, and a coarse graining of $N=100$ and $N=1000$ is considered. If each blade of a three-blade neutron interferometer contains $N$-planes, then the output on the third blade has $3N$-nodes, and hence the beam size increases at each blade. Note however that the plotted profiles in the figure are normalized in width. We find that the simulations are in agreement with the profiles generated by the application of the standard theory of DD to a NI \cite{Sam}, which are also shown in Fig.~\ref{Ubonse} for comparison.
 %%%%===============================================================================================================
\begin{figure}
\center
\includegraphics[scale=.35]{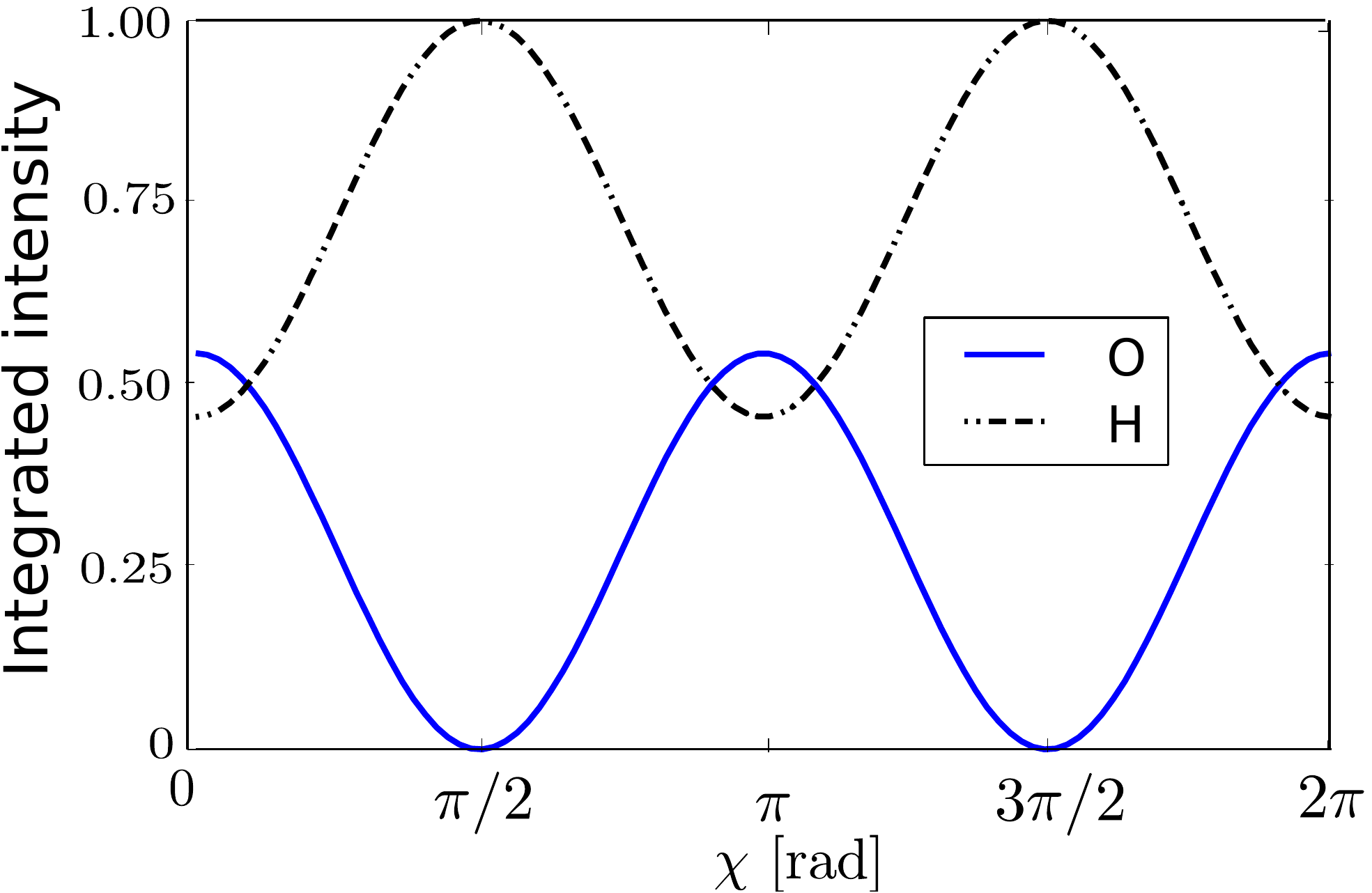}
\caption{The intensity of the O and H -beams as a function of the phase beteween the two NI paths $\chi$. The simulation is done for $N=100$ planes in each blade where the unitary operator of a single node is $U_{j,0,0,\pi/4}$.} \label{FOH}
\end{figure}
%%%%===============================================================================================================

\subsection{{Output Intensities}}

%---------------beginning of add------------------
The quantum information model is applied to simulate the output integrated intensities of a three blade neutron interferometer. Consider projectors onto the upward propagating beam ($O$) and the downward propagating beam ($H$)
 \[P_O=\sum_j\ket{a_j}\bra{a_j}\qquad P_H=\sum_j\ket{b_j}\bra{b_j}
 \]
 If the phase difference  between the two paths is $\chi$ (phase operator $U_z(\chi)=\exp[i\chi\sum_j(\ket{a_j}\bra{a_j}-\ket{b_j}\bra{b_j})/2]$) and denoting the operator of the first and last blade as $U_B$ and the middle blade as $U_M$, the wavefunction at the output is
\begin{align}
\ket{\Psi}&=\ket{\Psi^O}+\ket{\Psi^H}=U_BU_MU_z(\chi)U_B\ket{\Psi_{0}},
\end{align}
where the $O$ and $H$ components at the output can be written as
\begin{align}
\ket{\Psi^O}&=\sum_j\psi_{j,\bar{r}rt}\left(e^{-i\chi/2}+e^{i\chi/2}\right)\ket{a_j}\\
\ket{\Psi^H}&=\sum_j\left(e^{-i\chi/2}\psi_{j,\bar{t}rt}
+e^{i\chi/2}\psi_{j,r\bar{r}r}\right)\ket{b_j}.
\end{align}
where $\psi_{j,\bar{r}rt}=\bra{a_j}U_BP_HU_MP_OU_B\ket{\Psi_0}$, and $\psi_{j,\bar{t}rt}$ and $\psi_{j,r\bar{r}r}$ are  similarly obtained.
We note that the output from the QI model can be compared to that of the standard DD in Eqs.~(\ref{Eq:3bladeO},\ref{Eq:3bladeH}).

The sum intensity for $O$ and $H$ beams as a function of the phase difference between the paths is given by
\bea
%I_O&=&2\sum_i|(\psi_{\bar{r}rt})_i|^2(1+\cos\varphi)\\
I_O={\cal{A}}(1+\cos\chi),\quad
 %I_H&=&\sum_i\left[|(\psi_{\bar{r}rt})_i|^2+|(r\bar{r}r)_i|^2\right]
% \\& &-2\sum_i|(\psi_{\bar{r}rt})_i|^2\cos\varphi\\
I_H={\cal{B}}-{\cal{A}}\cos\chi
\eea
 with the coefficients,
 $ {\cal{A}}=2\sum_j|\psi_{j,\bar{r}rt}|^2$ and $  {\cal{B}}=\sum_j\left(|\psi_{j,\bar{t}rt}|^2+|\psi_{j,r\bar{r}r}|^2\right).$
%\bea\nonumber
% {\cal{A}}&=2\sum_j|\psi_{j,\bar{r}rt}|^2, \quad
% {\cal{B}}&=\sum_j\left(|\psi_{j,\bar{t}rt}|^2+|\psi_{j,r\bar{r}r}|^2\right).
%\eea
The intensities at the output of the NI for the $O$ and $H$ beams are presented in Fig.~\ref{FOH} for $N=100$ planes in each blade and the unitary $U_{j,0,0,\pi/4}$ at each node (note that $\theta=\pi/4$ corresponds to ($t_a=r_a=1/\sqrt{2}$). It is shown that the intensities of the output beams oscillate in a sinusoidal fashion and it can be noticed that the intensity of the $O$-beam has a minimum at zero. The well known asymmetry known from interferometry can be seen on the $H$-beam, where the intensity in this case never goes to zero.

\subsection{{Contrast}}
 A commonly used figure of merit in a NI is  the $contrast$ and it is obtained from the output intensities of the O- and H-beams. 
%\bea
%{\cal{V}}=\frac{\text{max}\{I(\varphi)\}-\text{min}\{I(\varphi)\}}{\text{max}\{I(\varphi)\}+\text{min}\{I(\varphi)\}}
%\eea
Under ideal conditions the contrast of the H-beam is 
\begin{align}\label{visibility}
{\cal{V}}_H&=\frac{\text{max}\{I_H(\chi)\}-\text{min}\{I_H(\chi)\}}{\text{max}\{I_H(\chi)\}+\text{min}\{I_H(\chi)\}}
=\frac{\cal{A}}{\cal{B}}
\end{align}
while that of the O-beam is always 1. In experiments, the contrast is always below 1 due to various reasons such as NI impurities, blade imperfections, external vibrations, and so on. 

\begin{figure}
\center
\includegraphics[scale=.45]{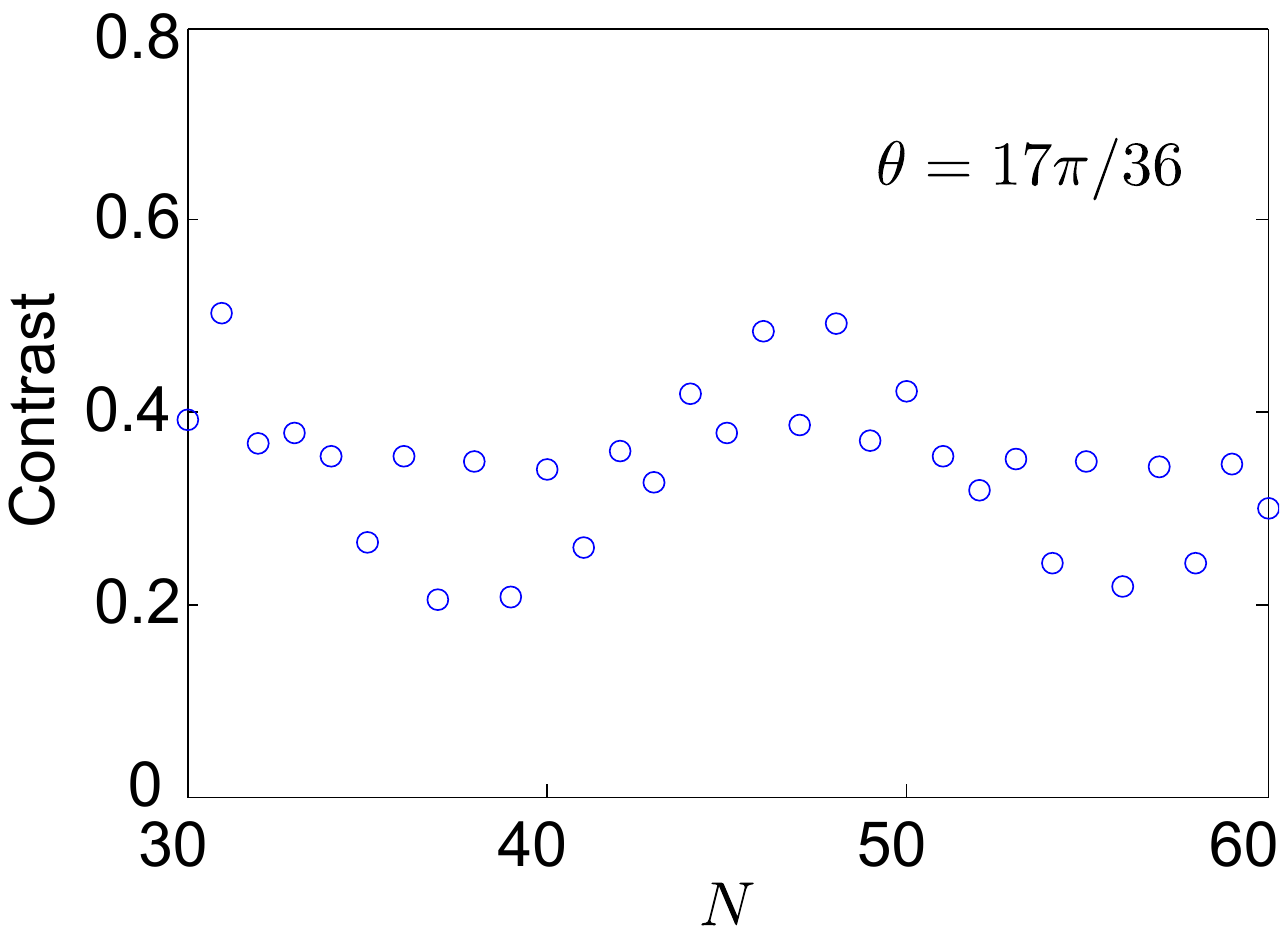}
\caption{The contrast of the H-beam for a three blade neutron interferometer as a function of the number of planes in each blade $N$. The unitary operator for each node is $U_{j,0,0,17\pi/36}$. To calculate the contrast the phase difference between the two paths is varied over a full cycle. The maximum and minimum are then extracted to get the contrast using Eq.~(\ref{visibility}).} \label{HContrast}
\end{figure}

In the QI model, to obtain the contrast for a fixed number of planes the phase difference between the two interferometer paths is varied over a full cycle. The maximum and minimum values are then extracted and the contrast is obtained using Eq.~(\ref{visibility}). The standard theory of DD predicts that the contrast on the H-beam converges to 0.39 for a three blade neutron interferometer \cite{Sam}. With the QI model, if $\theta$ is increased, the contrast on the $H$-beam reduces for a fixed number of planes. The 0.39 contrast obtained in the standard theory of DD could be reproduced in the QI model by using a suitable choice of $\theta$ that is close to $\pi/2$ (corresponding to $t\rightarrow 0, r\rightarrow 1$). Figure~\ref{HContrast} shows the contrast as a function of the number of planes obtained using the QI model where $\theta$ is fixed to $17\pi/36$ . In this limit the contrast converges to a similar value as predicted by the standard theory of DD.
%%%%===============================================================================================================
%%%%===============================================================================================================

%==============================================================
% Conclusion
%==============================================================
\section{Conclusion}

We have developed a quantum information model to study dynamical diffraction using unitary operators. This approach can help in the study of finer details of DD without reference to the parameters such as the particle wavelength, crystallographic orientation, and the interferometer geometry. To verify this quantum information model, it is used  to reproduce features of dynamical diffraction that have been studied before such as the intensities in the Borrmann triangle and Pendell{\"o}sung oscillation. A simple way to understand the variation of the contrast with  blade thickness is considered for a three blade neutron interferometer. The same process can be applied to other neutron interferometer geometries.

 The model could also be used to study concepts such as decoherence %during dynamical diffraction by a simple abstraction of the lattice to unitaries. These include 
 in cases where there is slight misalignment, missing atoms, isotopes, local absorption etc. In the case of missing atoms, the unitary of that particular node equals to an identity. In the case of misalignment, the parameters $\theta, \xi,$ and $ \zeta$ are different from the rest of the lattice.

Although we only considered elastic scattering, this model maybe applied to inelastic scattering by adding another degree of freedom.
 In general, one could entangle the particle’s momentum to other excitation (phonons, magnons, etc).  An example is DD of neutrons on a magnetic crystal where each incident neutron is in a superposition of spin up and spin down states. Due to the high magnetic field inside the magnetic crystal, spin up and spin down states will get different momenta and therefore will have different transmission and reflection coefficients. If the two degrees of freedoms (spin, momentum) are entangled, then by changing the entangled state from a symmetric in to an antisymmetric state, it is possible to mimic the dynamics of two non-interacting particles. The realization might be challenging in some cases as it requires a quadratically growing number of elements. However, this is outside of the scope of this paper.

\section{Acknowledgements}
This work was supported by the Canadian Excellence Research Chairs (CERC) program, the Natural Sciences and Engineering Research Council of Canada (NSERC) Discovery program, Collaborative Research and Training Experience (CREATE) program and the Canada First Research Excellence Fund (CFREF). The authors are grateful to Christopher Wood, and Thomas Alexander for useful discussions.

%==============================================================
%:References
%==============================================================
%\bibliographystyle{unsrt}
\bibliographystyle{apsrev4-1}

\bibliography{QWModelofDD}

\begin{thebibliography}{28}%
\makeatletter
\providecommand \@ifxundefined [1]{%
 \@ifx{#1\undefined}
}%
\providecommand \@ifnum [1]{%
 \ifnum #1\expandafter \@firstoftwo
 \else \expandafter \@secondoftwo
 \fi
}%
\providecommand \@ifx [1]{%
 \ifx #1\expandafter \@firstoftwo
 \else \expandafter \@secondoftwo
 \fi
}%
\providecommand \natexlab [1]{#1}%
\providecommand \enquote  [1]{``#1''}%
\providecommand \bibnamefont  [1]{#1}%
\providecommand \bibfnamefont [1]{#1}%
\providecommand \citenamefont [1]{#1}%
\providecommand \href@noop [0]{\@secondoftwo}%
\providecommand \href [0]{\begingroup \@sanitize@url \@href}%
\providecommand \@href[1]{\@@startlink{#1}\@@href}%
\providecommand \@@href[1]{\endgroup#1\@@endlink}%
\providecommand \@sanitize@url [0]{\catcode `\\12\catcode `\$12\catcode
  `\&12\catcode `\#12\catcode `\^12\catcode `\_12\catcode `\%12\relax}%
\providecommand \@@startlink[1]{}%
\providecommand \@@endlink[0]{}%
\providecommand \url  [0]{\begingroup\@sanitize@url \@url }%
\providecommand \@url [1]{\endgroup\@href {#1}{\urlprefix }}%
\providecommand \urlprefix  [0]{URL }%
\providecommand \Eprint [0]{\href }%
\providecommand \doibase [0]{http://dx.doi.org/}%
\providecommand \selectlanguage [0]{\@gobble}%
\providecommand \bibinfo  [0]{\@secondoftwo}%
\providecommand \bibfield  [0]{\@secondoftwo}%
\providecommand \translation [1]{[#1]}%
\providecommand \BibitemOpen [0]{}%
\providecommand \bibitemStop [0]{}%
\providecommand \bibitemNoStop [0]{.\EOS\space}%
\providecommand \EOS [0]{\spacefactor3000\relax}%
\providecommand \BibitemShut  [1]{\csname bibitem#1\endcsname}%
\let\auto@bib@innerbib\@empty
%</preamble>
\bibitem [{\citenamefont {Zachariasen}(1945)}]{Zachariasen67}%
  \BibitemOpen
  \bibfield  {author} {\bibinfo {author} {\bibfnamefont {W.~H.}\ \bibnamefont
  {Zachariasen}},\ }\href@noop {} {\emph {\bibinfo {title} {Theory of X-ray
  Diffraction in Crystals}}}\ (\bibinfo  {publisher} {Wiley},\ \bibinfo
  {address} {New York},\ \bibinfo {year} {1945})\BibitemShut {NoStop}%
\bibitem [{\citenamefont {Batterman}\ and\ \citenamefont
  {Cole}(1964)}]{Batterman&Cole}%
  \BibitemOpen
  \bibfield  {author} {\bibinfo {author} {\bibfnamefont {B.~W.}\ \bibnamefont
  {Batterman}}\ and\ \bibinfo {author} {\bibfnamefont {H.}~\bibnamefont
  {Cole}},\ }\href {\doibase 10.1103/RevModPhys.36.681} {\bibfield  {journal}
  {\bibinfo  {journal} {Rev. Mod. Phys.}\ }\textbf {\bibinfo {volume} {36}},\
  \bibinfo {pages} {681} (\bibinfo {year} {1964})}\BibitemShut {NoStop}%
\bibitem [{\citenamefont {Rauch}\ and\ \citenamefont
  {Petrascheck}(1978)}]{Rauch&Petrascheck78}%
  \BibitemOpen
  \bibfield  {author} {\bibinfo {author} {\bibfnamefont {H.}~\bibnamefont
  {Rauch}}\ and\ \bibinfo {author} {\bibfnamefont {D.}~\bibnamefont
  {Petrascheck}},\ }\href@noop {} {\bibfield  {journal} {\bibinfo  {journal}
  {Top. Curr. Phys.}\ }\textbf {\bibinfo {volume} {6}},\ \bibinfo {pages} {303}
  (\bibinfo {year} {1978})}\BibitemShut {NoStop}%
\bibitem [{\citenamefont {Sears}(1989)}]{Sears78}%
  \BibitemOpen
  \bibfield  {author} {\bibinfo {author} {\bibfnamefont {V.~F.}\ \bibnamefont
  {Sears}},\ }\href@noop {} {\emph {\bibinfo {title} {An Introduction to the
  Theory of Neutron Optical Phenomena and their Applications}}}\ (\bibinfo
  {publisher} {Oxford University Press},\ \bibinfo {address} {New York},\
  \bibinfo {year} {1989})\BibitemShut {NoStop}%
\bibitem [{\citenamefont {Authier}(2006)}]{Authier2006}%
  \BibitemOpen
  \bibfield  {author} {\bibinfo {author} {\bibfnamefont {A.}~\bibnamefont
  {Authier}},\ }\href@noop {} {\emph {\bibinfo {title} {Dynamical Theory of
  X-ray Diffraction}}}\ (\bibinfo  {publisher} {Oxford},\ \bibinfo {address}
  {Oxforn Univ Press},\ \bibinfo {year} {2006})\BibitemShut {NoStop}%
\bibitem [{\citenamefont {Darwin}(1914)}]{Darwin1914}%
  \BibitemOpen
  \bibfield  {author} {\bibinfo {author} {\bibfnamefont {C.~G.}\ \bibnamefont
  {Darwin}},\ }\href {\doibase 10.1080/14786440208635093} {\bibfield  {journal}
  {\bibinfo  {journal} {Philosophical Magazine Series 6}\ }\textbf {\bibinfo
  {volume} {27}},\ \bibinfo {pages} {315} (\bibinfo {year} {1914})}\BibitemShut
  {NoStop}%
\bibitem [{\citenamefont {James}(1963)}]{James1963}%
  \BibitemOpen
  \bibfield  {author} {\bibinfo {author} {\bibfnamefont {R.}~\bibnamefont
  {James}},\ }in\ \href {\doibase
  http://dx.doi.org/10.1016/S0081-1947(08)60592-5} {\emph {\bibinfo {booktitle}
  {The Dynamical Theory of X-Ray Diffraction}}},\ \bibinfo {series} {Solid
  State Physics}, Vol.~\bibinfo {volume} {15},\ \bibinfo {editor} {edited by\
  \bibinfo {editor} {\bibfnamefont {F.}~\bibnamefont {Seitz}}\ and\ \bibinfo
  {editor} {\bibfnamefont {D.}~\bibnamefont {Turnbull}}}\ (\bibinfo
  {publisher} {Academic Press},\ \bibinfo {address} {New York},\ \bibinfo
  {year} {1963})\ pp.\ \bibinfo {pages} {53 -- 220}\BibitemShut {NoStop}%
\bibitem [{\citenamefont {Reimer}(1984)}]{Reimer1984}%
  \BibitemOpen
  \bibfield  {author} {\bibinfo {author} {\bibfnamefont {L.}~\bibnamefont
  {Reimer}},\ }in\ \href {\doibase 10.1007/978-3-662-13553-2_7} {\emph
  {\bibinfo {booktitle} {Transmission Electron Microscopy}}},\ \bibinfo
  {series} {Springer Series in Optical Sciences}, Vol.~\bibinfo {volume} {36}\
  (\bibinfo  {publisher} {Springer Berlin Heidelberg},\ \bibinfo {year}
  {1984})\ pp.\ \bibinfo {pages} {259--313}\BibitemShut {NoStop}%
\bibitem [{\citenamefont {Oberthaler}\ \emph {et~al.}(1999)\citenamefont
  {Oberthaler}, \citenamefont {Abfalterer}, \citenamefont {Bernet},
  \citenamefont {Keller}, \citenamefont {Schmiedmayer},\ and\ \citenamefont
  {Zeilinger}}]{Oberthaler1999}%
  \BibitemOpen
  \bibfield  {author} {\bibinfo {author} {\bibfnamefont {M.~K.}\ \bibnamefont
  {Oberthaler}}, \bibinfo {author} {\bibfnamefont {R.}~\bibnamefont
  {Abfalterer}}, \bibinfo {author} {\bibfnamefont {S.}~\bibnamefont {Bernet}},
  \bibinfo {author} {\bibfnamefont {C.}~\bibnamefont {Keller}}, \bibinfo
  {author} {\bibfnamefont {J.}~\bibnamefont {Schmiedmayer}}, \ and\ \bibinfo
  {author} {\bibfnamefont {A.}~\bibnamefont {Zeilinger}},\ }\href {\doibase
  10.1103/PhysRevA.60.456} {\bibfield  {journal} {\bibinfo  {journal} {Phys.
  Rev. A}\ }\textbf {\bibinfo {volume} {60}},\ \bibinfo {pages} {456} (\bibinfo
  {year} {1999})}\BibitemShut {NoStop}%
\bibitem [{\citenamefont {Abov}\ \emph {et~al.}(2002)\citenamefont {Abov},
  \citenamefont {Elyutin},\ and\ \citenamefont {Tyulyusov}}]{Abov}%
  \BibitemOpen
  \bibfield  {author} {\bibinfo {author} {\bibfnamefont {Y.}~\bibnamefont
  {Abov}}, \bibinfo {author} {\bibfnamefont {N.}~\bibnamefont {Elyutin}}, \
  and\ \bibinfo {author} {\bibfnamefont {A.}~\bibnamefont {Tyulyusov}},\ }\href
  {\doibase 10.1134/1.1522085} {\bibfield  {journal} {\bibinfo  {journal}
  {Physics of Atomic Nuclei}\ }\textbf {\bibinfo {volume} {65}},\ \bibinfo
  {pages} {1933} (\bibinfo {year} {2002})}\BibitemShut {NoStop}%
\bibitem [{\citenamefont {Shull}(1968)}]{Shull63}%
  \BibitemOpen
  \bibfield  {author} {\bibinfo {author} {\bibfnamefont {C.~G.}\ \bibnamefont
  {Shull}},\ }\href {\doibase 10.1103/PhysRevLett.21.1585} {\bibfield
  {journal} {\bibinfo  {journal} {Phys. Rev. Lett.}\ }\textbf {\bibinfo
  {volume} {21}},\ \bibinfo {pages} {1585} (\bibinfo {year}
  {1968})}\BibitemShut {NoStop}%
\bibitem [{\citenamefont {Borrmann}(1950)}]{bormann}%
  \BibitemOpen
  \bibfield  {author} {\bibinfo {author} {\bibfnamefont {G.}~\bibnamefont
  {Borrmann}},\ }\href {\doibase 10.1007/BF01329828} {\bibfield  {journal}
  {\bibinfo  {journal} {Zeitschrift fur Physik}\ }\textbf {\bibinfo {volume}
  {127}},\ \bibinfo {pages} {297} (\bibinfo {year} {1950})}\BibitemShut
  {NoStop}%
\bibitem [{\citenamefont {Zeilinger}(1981)}]{Zeilinger81}%
  \BibitemOpen
  \bibfield  {author} {\bibinfo {author} {\bibfnamefont {A.}~\bibnamefont
  {Zeilinger}},\ }\href@noop {} {\bibfield  {journal} {\bibinfo  {journal}
  {American J. Phys.}\ }\textbf {\bibinfo {volume} {49}},\ \bibinfo {pages}
  {882} (\bibinfo {year} {1981})}\BibitemShut {NoStop}%
\bibitem [{\citenamefont {Lemmel}(2013)}]{Lemmel2013}%
  \BibitemOpen
  \bibfield  {author} {\bibinfo {author} {\bibfnamefont {H.}~\bibnamefont
  {Lemmel}},\ }\href {\doibase 10.1107/S0108767313014293} {\bibfield  {journal}
  {\bibinfo  {journal} {Acta Crystallographica Section A}\ }\textbf {\bibinfo
  {volume} {69}},\ \bibinfo {pages} {459} (\bibinfo {year} {2013})}\BibitemShut
  {NoStop}%
\bibitem [{\citenamefont {Lemmel}(2007)}]{Lemmel2007}%
  \BibitemOpen
  \bibfield  {author} {\bibinfo {author} {\bibfnamefont {H.}~\bibnamefont
  {Lemmel}},\ }\href {\doibase 10.1103/PhysRevB.76.144305} {\bibfield
  {journal} {\bibinfo  {journal} {Phys. Rev. B}\ }\textbf {\bibinfo {volume}
  {76}},\ \bibinfo {pages} {144305} (\bibinfo {year} {2007})}\BibitemShut
  {NoStop}%
\bibitem [{\citenamefont {Rauch}\ and\ \citenamefont {Werner}(2000)}]{Sam}%
  \BibitemOpen
  \bibfield  {author} {\bibinfo {author} {\bibfnamefont {H.}~\bibnamefont
  {Rauch}}\ and\ \bibinfo {author} {\bibfnamefont {S.}~\bibnamefont {Werner}},\
  }\href@noop {} {\emph {\bibinfo {title} {Neutron Interferometry: Lesson in
  Experimental Quantum Mechanics}}}\ (\bibinfo  {publisher} {Claredon Press},\
  \bibinfo {address} {Oxford},\ \bibinfo {year} {2000})\BibitemShut {NoStop}%
\bibitem [{\citenamefont {Utsuro}\ and\ \citenamefont
  {Ignatovich}(2012)}]{Vladimir}%
  \BibitemOpen
  \bibfield  {author} {\bibinfo {author} {\bibfnamefont {M.}~\bibnamefont
  {Utsuro}}\ and\ \bibinfo {author} {\bibfnamefont {V.~K.}\ \bibnamefont
  {Ignatovich}},\ }\href@noop {} {\emph {\bibinfo {title} {Handbook of Neutron
  Optics}}}\ (\bibinfo  {publisher} {Wiley-VCH},\ \bibinfo {year}
  {2012})\BibitemShut {NoStop}%
\bibitem [{\citenamefont {Bouwmeester}\ \emph {et~al.}(1999)\citenamefont
  {Bouwmeester}, \citenamefont {Marzoli}, \citenamefont {Karman}, \citenamefont
  {Schleich},\ and\ \citenamefont {Woerdman}}]{Galton}%
  \BibitemOpen
  \bibfield  {author} {\bibinfo {author} {\bibfnamefont {D.}~\bibnamefont
  {Bouwmeester}}, \bibinfo {author} {\bibfnamefont {I.}~\bibnamefont
  {Marzoli}}, \bibinfo {author} {\bibfnamefont {G.~P.}\ \bibnamefont {Karman}},
  \bibinfo {author} {\bibfnamefont {W.}~\bibnamefont {Schleich}}, \ and\
  \bibinfo {author} {\bibfnamefont {J.~P.}\ \bibnamefont {Woerdman}},\ }\href
  {\doibase 10.1103/PhysRevA.61.013410} {\bibfield  {journal} {\bibinfo
  {journal} {Phys. Rev. A}\ }\textbf {\bibinfo {volume} {61}},\ \bibinfo
  {pages} {013410} (\bibinfo {year} {1999})}\BibitemShut {NoStop}%
\bibitem [{\citenamefont {Aharonov}\ \emph {et~al.}(1993)\citenamefont
  {Aharonov}, \citenamefont {Davidovich},\ and\ \citenamefont {Zagury}}]{qw}%
  \BibitemOpen
  \bibfield  {author} {\bibinfo {author} {\bibfnamefont {Y.}~\bibnamefont
  {Aharonov}}, \bibinfo {author} {\bibfnamefont {L.}~\bibnamefont
  {Davidovich}}, \ and\ \bibinfo {author} {\bibfnamefont {N.}~\bibnamefont
  {Zagury}},\ }\href {\doibase 10.1103/PhysRevA.48.1687} {\bibfield  {journal}
  {\bibinfo  {journal} {Phys. Rev. A}\ }\textbf {\bibinfo {volume} {48}},\
  \bibinfo {pages} {1687} (\bibinfo {year} {1993})}\BibitemShut {NoStop}%
\bibitem [{\citenamefont {Venegas-Andraca}(2012)}]{QWalk}%
  \BibitemOpen
  \bibfield  {author} {\bibinfo {author} {\bibfnamefont {S.}~\bibnamefont
  {Venegas-Andraca}},\ }\href {\doibase 10.1007/s11128-012-0432-5} {\bibfield
  {journal} {\bibinfo  {journal} {Quantum Information Processing}\ }\textbf
  {\bibinfo {volume} {11}},\ \bibinfo {pages} {1015} (\bibinfo {year}
  {2012})}\BibitemShut {NoStop}%
\bibitem [{\citenamefont {Darwin}(1922)}]{Darwin1922}%
  \BibitemOpen
  \bibfield  {author} {\bibinfo {author} {\bibfnamefont {C.~G.}\ \bibnamefont
  {Darwin}},\ }\href {\doibase 10.1080/14786442208633940} {\bibfield  {journal}
  {\bibinfo  {journal} {Philosophical Magazine Series 6}\ }\textbf {\bibinfo
  {volume} {43}},\ \bibinfo {pages} {800} (\bibinfo {year} {1922})}\BibitemShut
  {NoStop}%
\bibitem [{\citenamefont {Ambartsumyan}(1981)}]{Ambartsumyan81}%
  \BibitemOpen
  \bibfield  {author} {\bibinfo {author} {\bibfnamefont {V.~A.}\ \bibnamefont
  {Ambartsumyan}},\ }\href@noop {} {\emph {\bibinfo {title} {Proceedings of the
  All Union Symposium Confined 40th Anniversary of Leading of Invariance
  Principle in RadiativeTransport Theory}}}\ (\bibinfo  {publisher} {Akad. Nauk
  Arm. SSR},\ \bibinfo {address} {Yerevan},\ \bibinfo {year}
  {1981})\BibitemShut {NoStop}%
\bibitem [{\citenamefont {Shull}(1973)}]{Shull73a}%
  \BibitemOpen
  \bibfield  {author} {\bibinfo {author} {\bibfnamefont {C.~G.}\ \bibnamefont
  {Shull}},\ }\href {\doibase 10.1107/S0021889873008654} {\bibfield  {journal}
  {\bibinfo  {journal} {Journal of Applied Crystallography}\ }\textbf {\bibinfo
  {volume} {6}},\ \bibinfo {pages} {257} (\bibinfo {year} {1973})}\BibitemShut
  {NoStop}%
\bibitem [{\citenamefont {Sippel}\ \emph {et~al.}(1965)\citenamefont {Sippel},
  \citenamefont {Kleinstuck},\ and\ \citenamefont {Schulze}}]{Sippel1965}%
  \BibitemOpen
  \bibfield  {author} {\bibinfo {author} {\bibfnamefont {D.}~\bibnamefont
  {Sippel}}, \bibinfo {author} {\bibfnamefont {K.}~\bibnamefont {Kleinstuck}},
  \ and\ \bibinfo {author} {\bibfnamefont {G.~E.~R.}\ \bibnamefont {Schulze}},\
  }\href {\doibase http://dx.doi.org/10.1016/0031-9163(65)90570-6} {\bibfield
  {journal} {\bibinfo  {journal} {Physics Letters}\ }\textbf {\bibinfo {volume}
  {14}},\ \bibinfo {pages} {174 } (\bibinfo {year} {1965})}\BibitemShut
  {NoStop}%
\bibitem [{\citenamefont {Somenkov}\ \emph {et~al.}(1978)\citenamefont
  {Somenkov}, \citenamefont {Shilstein}, \citenamefont {Belova},\ and\
  \citenamefont {Utemisov}}]{Somenkov1978}%
  \BibitemOpen
  \bibfield  {author} {\bibinfo {author} {\bibfnamefont {V.~A.}\ \bibnamefont
  {Somenkov}}, \bibinfo {author} {\bibfnamefont {S.~S.}\ \bibnamefont
  {Shilstein}}, \bibinfo {author} {\bibfnamefont {N.~E.}\ \bibnamefont
  {Belova}}, \ and\ \bibinfo {author} {\bibfnamefont {K.}~\bibnamefont
  {Utemisov}},\ }\href {\doibase
  http://dx.doi.org/10.1016/0038-1098(78)91497-7} {\bibfield  {journal}
  {\bibinfo  {journal} {Solid State Communications}\ }\textbf {\bibinfo
  {volume} {25}},\ \bibinfo {pages} {593} (\bibinfo {year} {1978})}\BibitemShut
  {NoStop}%
\bibitem [{\citenamefont {Sudo}(2006)}]{Sudo06}%
  \BibitemOpen
  \bibfield  {author} {\bibinfo {author} {\bibfnamefont {M.}~\bibnamefont
  {Sudo}},\ }\href@noop {} {\emph {\bibinfo {title} {Quantum Interferometry in
  Phase Space}}}\ (\bibinfo  {publisher} {Springer},\ \bibinfo {year}
  {2006})\BibitemShut {NoStop}%
\bibitem [{\citenamefont {Pushin}\ \emph {et~al.}(2011)\citenamefont {Pushin},
  \citenamefont {Huber}, \citenamefont {Arif},\ and\ \citenamefont
  {Cory}}]{Dima2011}%
  \BibitemOpen
  \bibfield  {author} {\bibinfo {author} {\bibfnamefont {D.~A.}\ \bibnamefont
  {Pushin}}, \bibinfo {author} {\bibfnamefont {M.~G.}\ \bibnamefont {Huber}},
  \bibinfo {author} {\bibfnamefont {M.}~\bibnamefont {Arif}}, \ and\ \bibinfo
  {author} {\bibfnamefont {D.~G.}\ \bibnamefont {Cory}},\ }\href {\doibase
  10.1103/PhysRevLett.107.150401} {\bibfield  {journal} {\bibinfo  {journal}
  {Phys. Rev. Lett.}\ }\textbf {\bibinfo {volume} {107}},\ \bibinfo {pages}
  {150401} (\bibinfo {year} {2011})}\BibitemShut {NoStop}%
\bibitem [{\citenamefont {Bonse}\ and\ \citenamefont {Rauch}(1979)}]{Ubonse}%
  \BibitemOpen
  \bibfield  {author} {\bibinfo {author} {\bibfnamefont {U.}~\bibnamefont
  {Bonse}}\ and\ \bibinfo {author} {\bibfnamefont {H.}~\bibnamefont {Rauch}},\
  }\href@noop {} {\emph {\bibinfo {title} {Neutron Interferometry}}}\ (\bibinfo
   {publisher} {Oxford University Press},\ \bibinfo {address} {Oxford, UK},\
  \bibinfo {year} {1979})\BibitemShut {NoStop}%
\end{thebibliography}%

\end{document}